\documentclass{emulateapj}
\usepackage{float}
\usepackage{amsmath}

\begin{document}

\title{The Scaling of Stellar Mass and Central Stellar Velocity Dispersion for Quiescent galaxies at $\lowercase{z}<0.7$}

\author{H. Jabran Zahid$^{1}$\footnote{email: zahid@cfa.harvard.edu}, Margaret J. Geller$^{1}$ Daniel G. Fabricant$^{1}$ \& Ho Seong Hwang$^{2}$}
\affil{$^{1}$Smithsonian Astrophysical Observatory, Harvard-Smithsonian Center for Astrophysics - 60 Garden Street, Cambridge, MA 02138}
\affil{$^{2}$School of Physics, Korea Institute for Advanced Study, 85 Hoegiro, Dongdaemun-gu, Seoul 02455, Republic of Korea}

\def\mean#1{\left< #1 \right>}

\begin{abstract}

We examine the relation between stellar mass and central stellar velocity dispersion$-$the $M_\ast \sigma$ relation$-$for massive quiescent galaxies at $z<0.7$. We measure the local relation from the Sloan Digital Sky Survey and the intermediate redshift relation from the Smithsonian Hectospec Lensing Survey. Both samples are highly complete ($>85\%$) and we consistently measure the stellar mass and velocity dispersion for the two samples. The $M_\ast \sigma$ relation and its scatter are independent of redshift with $\sigma \propto M_\ast^{0.3}$ for $M_\ast \gtrsim 10^{10.3} M_\odot$. The measured slope of the $M_\ast \sigma$ relation is the same as the scaling between the total halo mass and the dark matter halo velocity dispersion obtained by N-body simulations. This consistency suggests that massive quiescent galaxies are virialized systems where the central dark matter concentration is either a constant or negligible fraction of the stellar mass. The relation between the total galaxy mass (stellar + dark matter) and the central stellar velocity dispersion is consistent with the observed relation between the total mass of a galaxy cluster and the velocity dispersion of the cluster members. This result suggests that the central stellar velocity dispersion is directly proportional to the velocity dispersion of the dark matter halo. Thus the central stellar velocity dispersion is a fundamental, directly observable property of galaxies that may robustly connect galaxies to dark matter halos in N-body simulations. To interpret the results further in the context of $\Lambda$CDM, it would be useful to analyze the relationship between the velocity dispersion of stellar particles and the velocity dispersion characterizing their dark matter halos in high-resolution cosmological hydrodynamic simulations.

\end{abstract}
\keywords{galaxies: evolution $-$ galaxies: high-redshift $-$ galaxies: formation $-$ galaxies: structure}

\section{Introduction}

In the standard model of cosmology the vast majority of matter in the universe ($\sim84\%$) is dark \citep{Planck2015}. The large scale structure of the universe develops as gravity acts on small density fluctuations in the virtually uniform dark matter dominated initial matter distribution. Gravity forms and shapes dark matter into nearly spherical halos. Baryons accrete onto dark matter halos and subsequently cool and condense to form galaxies. Dark matter halos are more extended and substantially more massive than the galaxies which form and evolve at their centers. Within the hierarchical formation paradigm, large scale structure formation and galaxy evolution are primarily driven by the accretion of dark matter and by halo mergers. Dark matter can not be directly observed and thus a central issue for cosmology is what observable property is the best proxy for connecting galaxies$-$the visible tracers of the matter distribution$-$to the dark matter distribution. 

Within the broader cosmological context, connections among the observed properties of galaxies should elucidate the physical processes governing galaxy formation and evolution. Statistical analyses of the galaxy population have established that the principal observable galaxy properties are all correlated, though the fundamental parameter in these correlations remains uncertain \citep{Disney2008}. Several recent studies suggest that either stellar mass or stellar velocity dispersion is the fundamental parameter characterizing galaxies and their dark matter halos \citep{More2011, Wake2012a, Wake2012b, Li2013, vanUitert2013, Bogdan2015}.

The stellar mass and velocity dispersion are governed by different physical processes, but both are intimately related to properties of the dark matter halo. Stellar mass and velocity dispersion are strongly correlated making it difficult to determine which of these two parameters is fundamental. The stellar mass is an integral over the star formation history and the end product of the complex baryonic processes governing galaxy formation and evolution. In contrast, the velocity dispersion is a measure of the stellar kinematics and is directly related to the gravitational potential of the system.   

The correlation between luminosity and velocity dispersion in elliptical galaxies is well established. Based on observations of 25 galaxies, \citet{Faber1976} find a power law relation between velocity dispersion and luminosity and conclude that the total mass is the most fundamental property of elliptical galaxies. The stellar mass-to-light ratio for elliptical galaxies does not vary significantly; a relation between stellar mass and velocity dispersion directly follows from the \citet{Faber1976} result. Many subsequent studies based on larger samples and/or spatially resolved spectroscopy confirm a power law relation between stellar mass and velocity dispersion {over most of the stellar mass range explored in these studies} \citep[e.g.,][]{Hyde2009a, Cappellari2013b, Cappellari2016}.

The redshift evolution of the relation between stellar mass and velocity dispersion and the scatter around the relation provide further constraints for models of galaxies. \citet{Belli2014} show that the relation between stellar mass and velocity dispersion at $0.9<z<1.6$ is offset from the local relation. \citet{Belli2014} attribute this offset to the smaller sizes of galaxies at higher redshift. This interpretation is consistent with the fact that the intrinsic scatter in the relation between luminosity and stellar mass is strongly correlated with size$-$the so-called fundamental plane \citep{Djorgovski1987, Dressler1987}; this relation also exists when luminosity is replaced by stellar mass \citep{Hyde2009b}. The stellar mass fundamental plane does not appear to evolve strongly for $z<0.6$ \citep{Zahid2016} but may at higher redshifts \citep{Bezanson2013}. Thus, the smaller sizes of galaxies at high redshift may explain their larger velocity dispersions at a fixed stellar mass. 

\citet{Shu2012} analyze the velocity dispersion distribution of a large sample of luminous red galaxies observed as part of the Baryon Oscillation Spectroscopic Survey (BOSS). Due to the limiting signal-to-noise of their observations, \citet{Shu2012} measure the intrinsic scatter by a ``Bayesian stacking" technique rather than from examination of the distribution of individual galaxy velocity dispersions. They report that the intrinsic scatter in velocity dispersions at a fixed stellar mass increases as a function of redshift, though the evolution is small for $z\lesssim0.6$ \citep[see Figure 11 in][]{Shu2012}. They conclude that the increased scatter indicates greater diversity in the galaxy population at early times. {In a more recent analysis of the BOSS sample, \citet{Montero-Dorta2016} find that the slope and scatter in the relation between luminosity and velocity dispersion for high-mass red sequence galaxies does not evolve significantly between $0.5<z<0.7$. Thus, studies of large samples of red galaxies from BOSS indicate little evolution in the relation between velocity dispersions and stellar mass or luminosity.}


Here we analyze a sample of 4585 galaxies with individual stellar mass and velocity dispersion measurements at $z<0.7$ to examine the stellar mass-velocity dispersion ($M_\ast \sigma$) relation and its dependence on redshift. The intermediate redshift range we probe connects SDSS observations to the higher redshift observations of \citet{Belli2014} and our analysis of the scatter around the $M_\ast \sigma$ relation is complementary to \citet{Shu2012}; we measure a velocity dispersion for each galaxy.  In Section 2 we describe the sample and present the $M_\ast \sigma$ relation in Section 3. We analyze the scatter in the $M_\ast \sigma$ relation in Section 4 and highlight important systematic issues in Section 5. We discuss the results in Section 6 and summarize and conclude in Section 7. We adopt the standard cosmology $(H_{0}, \Omega_{m}, \Omega_{\Lambda}) = (70$ km s$^{-1}$ Mpc$^{-1}$, 0.3, 0.7) throughout.

\section{Observations, Methods and Sample Selection}

\begin{figure*}
\begin{center}
\includegraphics[width = 2 \columnwidth]{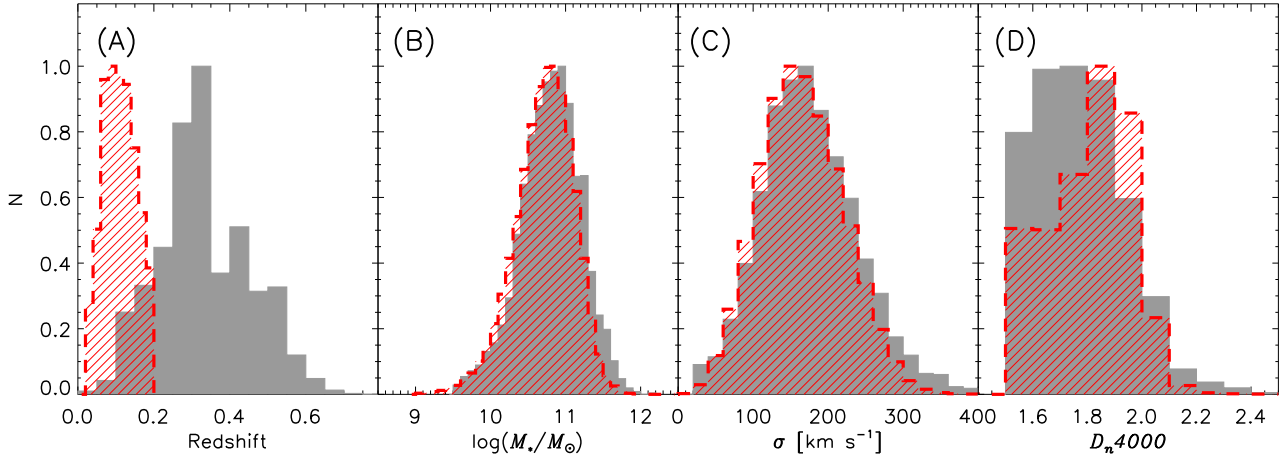}
\end{center}
\caption{Histogram of the (A) redshift, (B) stellar mass, (C) velocity dispersion and (D) $D_n4000$ index for the SDSS (red hashed) and SHELS (gray) samples. }
\label{fig:hist}
\end{figure*}

\subsection{Survey Data}

We analyze a local sample from the Sloan Digital Sky Survey DR12\footnote{http://www.sdss.org/dr12/} \citep{Alam2015}. We restrict the analysis to the Main Galaxy Sample of $\sim900,000$ galaxies with $r<17.8$ observed over $\sim10,000$ deg$^{2}$ in the redshift range $0\lesssim z \lesssim 0.3$ \citep{York2000}. The nominal spectral range of the SDSS observations is $3800 - 9200 \mathrm{\AA}$ at a resolution of $R\sim1500$ at $5000\mathrm{\AA}$ \citep{Smee2013}. We use the $ugriz$ model magnitudes from the SDSS imaging data \citep{Stoughton2002, Doi2010}. 

The intermediate redshift sample is from the Smithsonian Hectospec Lensing Survey \citep[SHELS;][]{Geller2005, Geller2014, Geller2016}. The redshift survey covers two 4 deg$^{2}$ fields (F1 and F2) of the Deep Lensing Survey \citep[DLS;][]{Wittman2002}. Here we analyze the F2 field only and refer to this sample as the SHELS sample throughout this work. Redshifts, stellar masses and $D_n4000$ indices for galaxies in F2 are published in \citet{Geller2014}. The velocity dispersion measurements for F1 and F2 are forthcoming (Geller et al., in prep). The survey is $\gtrsim90\%$ complete at $R<20.6$ ($r\lesssim20.9$) and consists of $\sim13,300$ galaxies observed in the redshift range of $0<z<0.7$. This level of spectroscopic completeness is comparable to the SDSS Main Galaxy Sample at $r<17.8$ \citep{Strauss2002}. The spectra are obtained with Hectospec, a 300 fiber optical spectrograph on the 6.5m MMT \citep{Fabricant2005}. The nominal spectral range of the observations is $3700-9100 \mathrm{\AA}$ at a resolution of $R\sim1000$ at $5000\mathrm{\AA}$. We use the SDSS $ugriz$ model magnitudes from the imaging pipeline \citep{Stoughton2002}.

\subsection{Stellar Mass}

To determine stellar masses we estimate the mass-to-light (M/L) ratio by $\chi^2$ fitting synthetic spectral energy distributions (SEDs) to the observed photometry. Stellar masses derived from SED fitting carry absolute uncertainties of $\sim0.3$ dex \citep{Conroy2009a}. These uncertainties arise from uncertainties in the star formation history (SFH), metallicity, dust extinction, stellar templates and IMF adopted to fit the SED. Our analysis relies on the \emph{relative} accuracy of stellar mass estimates. To mitigate relative offsets, we calculate all stellar masses from $ugriz$ SDSS model magnitudes using a consistent implementation of the {\sc Lephare}\footnote{http://www.cfht.hawaii.edu/$\sim$arnouts/LEPHARE/lephare.html} fitting code \citep{Arnouts1999, Ilbert2006b}. By comparing dispersion in two independent SED fitting methods, we estimate a $\sim0.1$ dex relative uncertainty in stellar masses calculated using {\sc Lephare} \citep[e.g.,][]{Zahid2014b}. This dispersion is dominated by the observational uncertainties; {this does not account for systematic uncertainties in the photometry.}

We fit the observed SED with {\sc Lephare} using the stellar population synthesis models of \citet{Bruzual2003} and the \citet{Chabrier2003} initial mass function (IMF). The procedure explicitly assumes a universal IMF. The models have two metallicities and exponentially declining SFHs (star formation rate $\propto e^{-t/\tau}$) with e-folding times of $\tau = 0.1,0.3,1,2,3,5,10,15$ and $30$ Gyr. We generate synthetic SEDs from these models by varying the extinction and stellar population age. We adopt the \citet{Calzetti2000} extinction law and allow $E(B-V)$ to range from 0 to 0.6. The stellar population ages range between 0.01 and 13 Gyr. Each synthetic SED is normalized to solar luminosity and the scale factor between the observed and synthetic SED is the stellar mass. This procedure yields a distribution for the best-fit stellar mass and we adopt the median of this distribution\footnote{{Our stellar mass estimates are systematically smaller by 0.13 dex compared to the stellar mass estimates of the Portsmouth group (medianpdf in the Passive Kroupa stellar mass catalog).}} Figure \ref{fig:hist}B shows the resulting stellar mass distributions for galaxies in the SDSS and SHELS samples (quiescent galaxies only; see Section 2.5 for detailed sample selection).

\subsection{Velocity Dispersion}

The line-of-sight (LOS) velocity dispersions for the SDSS and SHELS samples are measured from stellar absorption lines observed through circular fiber apertures centered on each galaxy. The velocity dispersion is measured in km s$^{-1}$. We refer to the central LOS stellar velocity dispersion as the velocity dispersion and denote it with the symbol $\sigma$. We denote the observational uncertainty in $\sigma$ by the symbol $\Delta$.

\citet{Thomas2013} measure velocity dispersions for SDSS galaxies from spectra observed through 3" fiber apertures. They use the Penalized Pixel-Fitting (p{\sc PXF}) code \citep{Cappellari2004} and the \citet{Maraston2011} stellar population templates based on the MILES stellar library \citep{Sanchez-Blazquez2006}. The templates are matched to the SDSS resolution and are parameterized by convolution with $\sigma$. The best-fit $\sigma$ is determined by minimizing the $\chi^2$ between the model and observed spectrum in the rest-frame wavelength range of $4500-6500\mathrm{\AA}$. The spectral resolution of SDSS limits reliable estimates of $\sigma$ to $\gtrsim60$ km s$^{-1}$.

We measure $\sigma$ for the SHELS sample from spectra observed through the 1''\!\!.5 fiber aperture of Hectospec \citep[for details see][]{Fabricant2013}. To measure $\sigma$ we use the University of Lyon Spectroscopic analysis Software \citep[ULySS;][]{Koleva2009}. Stellar population templates are calculated with the {\sc PEGASE-HR} code \citep{LeBorgne2004} and the MILES stellar library. The templates are matched to the Hectospec resolution, convolved with varying velocity dispersions and are parameterized by age and metallicity. The best-fit age, metallicity and $\sigma$ are determined from a $\chi^2$ fit of the convolved templates to each spectrum. The fit is limited to the rest-frame spectral range of $4100-5500\mathrm{\AA}$. \citet{Fabricant2013} find this spectral range minimizes $\Delta$ and provides the most stable results for Hectospec data. This rest-frame spectral range is fully accessible with Hectospec for $z<0.65$. Less than 1\% of galaxies in the SHELS sample have redshifts beyond this limit. The spectral resolution of Hectospec limits reliable estimates of the velocity dispersion to $\gtrsim90$ km s$^{-1}$. $\Delta$ is determined from the observational uncertainty in the spectrum. Based on a comparison of independent repeat observations of the same galaxy, \citet{Fabricant2013} determine that the $\Delta$ returned by the procedure is robust for galaxies with observational errors $<30$ km s$^{-1}$. For galaxies with multiple observations, we take the $\Delta$-weighted $\sigma$ as our measure of the velocity dispersion.

\begin{figure}
\begin{center}
\includegraphics[width = \columnwidth]{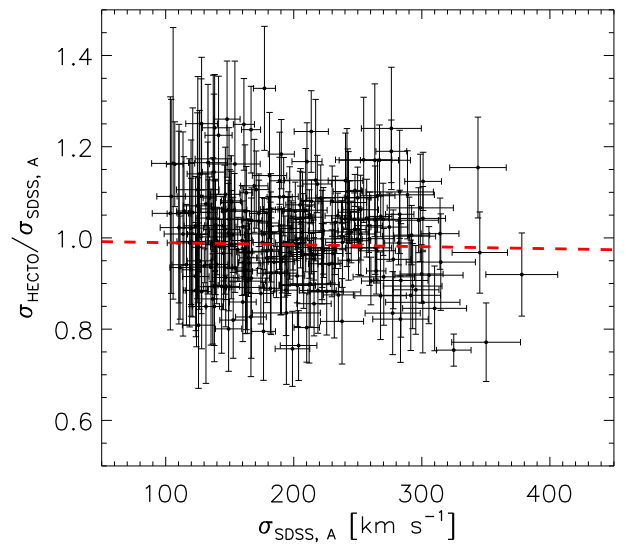}
\end{center}
\caption{Comparison of the aperture corrected velocity dispersion (see Equation \ref{eq:apcorr}) measured for the same galaxies with Hectospec and SDSS. The red line is the best-fit linear relation (Equation \ref{eq:vdisp_comp}).}
\label{fig:vdisp_comp}
\end{figure}

We cross-calibrate the SDSS and Hectospec measurements of $\sigma$ by comparing velocity dispersions measured for the same object. We match galaxies with SDSS and Hectospec spectra in the SHELS field by requiring $<$0''.5 separation. {We restrict the comparison to galaxies with $100<\sigma<450$ km s$^{-1}$ and  $\Delta<30$ km s$^{-1}$. We remove 3 outliers. The final sample consists of 247 galaxies observed in both SDSS and SHELS.}

We compare the samples to derive the cross-calibration. The largest systematic correction required is for the differing aperture sizes, thus we derive this correction first. We then compare the data to determine any residual systematic difference.

We parameterize the aperture correction as
\begin{equation}
\frac{\sigma_{\mathrm{SDSS}}}{\sigma_{\mathrm{HECTO}}} = \left( \frac{R_{\mathrm{SDSS}}}{R_{\mathrm{HECTO}}} \right)^\beta, 
\label{eq:apcorr}
\end{equation}
where $R_{\mathrm{SDSS}} = 1''\!\!.5$ and $R_{\mathrm{HECTO}} = 0''\!\!.75$ are the fiber aperture {radii}. We find $\beta = -0.033 \pm  0.011$ minimizes the difference between the SDSS and SHELS measurements. This value is consistent with \citet{Cappellari2006} who measure $\beta = -0.066 \pm0.035$ from integral field observations \citep[see also][]{Jorgensen1995, Mehlert2003}. We apply the aperture correction to the SDSS measurements and examine the residuals between the Hectospec and SDSS velocity dispersions. 

Figure \ref{fig:vdisp_comp} shows a comparison of the two velocity dispersion measurements {after the SDSS velocity dispersions are corrected to the Hectospec aperture}. We fit the residual systematic difference accounting for errors in both the coordinates using \emph{fitexy.pro} in the {\sc IDL Astronomy Library}. The best-fit is
\begin{equation}
\frac{\sigma_{\mathrm{HECTO, A}}}{\sigma_{\mathrm{SDSS, A}}} =  (1.003 \pm 0.005)  -  (4 \pm 9) \times 10^{-5} \sigma_{\mathrm{SDSS, A}}
\label{eq:vdisp_comp}
\end{equation}
The $A$ subscript indicates that the velocity dispersion is aperture corrected. {Within the errors, the zero point and slope are consistent with zero and unity, respectively. After applying the aperture correction, there is no residual statistically significant systematic difference in the SDSS and Hectospec measurements of velocity dispersion. The root-mean-square (RMS) of the difference in velocity dispersion is 21 km s$^{-1}$ which is slightly larger than the 18 km s$^{-1}$ expected from the observational uncertainties. Based on a detailed comparison of SDSS and Hectospec velocity dispersions, \citep{Fabricant2013} suggest that SDSS uncertainties may be underestimated by $\sqrt 2$. Multiplying the SDSS observational uncertainties by $\sqrt 2$ brings the difference expected from the observational uncertainties into agreement with the RMS difference in the two measurements.}

To compare the velocity dispersions across a broad range of stellar masses and redshifts, we correct all velocity dispersions to a fiducial physical aperture of 3 kpc using the correction in Equation \ref{eq:apcorr}. This fiducial physical aperture corresponds to the radius of a galaxy observed through a 1''\!\!.5 fiber at the median redshift of the SHELS sample ($z\sim0.3$). {The aperture correction is small (between -0.01 to 0.02 dex) and none of our conclusions are affected if we use the aperture correction derived by \citet{Cappellari2006}.} Figure \ref{fig:hist}C shows the velocity dispersion distribution of the SDSS and SHELS galaxies.

\subsection{$D_{n}4000$ Index}

The $D_n4000$ index is the flux ratio between two spectral windows adjacent to the 4000$\mathrm{\AA}$ break \citep{Balogh1999}. We adopt the measurements from the MPA/JHU\footnote{http://wwwmpa.mpa-garching.mpg.de/SDSS/DR7/} group for galaxies in the SDSS \citep{Kauffmann2003a} and take $D_n4000$ indices for SHELS galaxies from \citet{Geller2014}. \citet{Fabricant2008} show that the relative fluxes measured for the same objects with SDSS and SHELS vary by $\sim 5\%$ over the spectral range $3850 - 8500 \mathrm{\AA}$. They find that the $D_n4000$ indices measured for this overlapping sample are consistent to within a few percent. This level of accuracy is sufficient for our analysis. 

The index distribution is bimodal \citep{Kauffmann2003a, Geller2014} and we use it to select quiescent galaxies containing older stellar populations where the stellar kinematics are typically dominated by random motions \citep[e.g.,][]{Brinchmann2004}. {The bimodality of the distribution is a consequence of the fact that the $D_n4000$ index is sensitive to the age of the stellar population and quiescent galaxies are dominated by older stars \citep[e.g.,][]{Kauffmann2003a}.} Figure \ref{fig:hist}D shows the $D_n4000$ distribution of the SHELS and SDSS samples. {The $D_n4000$ distribution of SHELS galaxies is shifted towards lower values, reflecting the younger age of the quiescent galaxy population at higher redshifts.}

\begin{figure*}
\begin{center}
\includegraphics[width =  2\columnwidth]{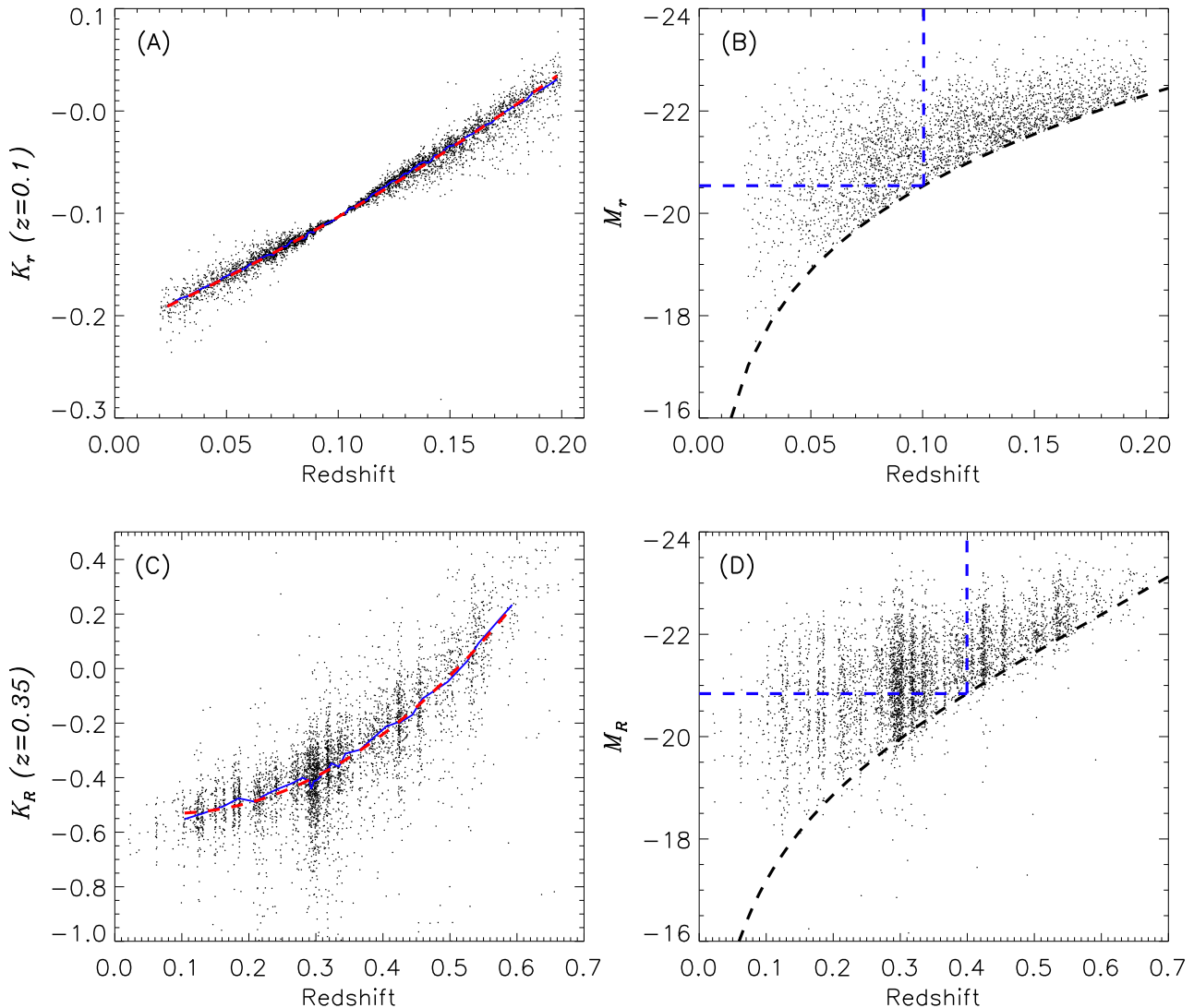}
\end{center}
\caption{(A) $r$-band $K$-correction for SDSS galaxies as a function of redshift. We show only a random subsample of 5000 galaxies for clarity. The blue line is the median $K$-correction in equally populated bins of redshift and the red line is a parabolic fit to the median correction. (B) Absolute $K$-corrected $r$-band magnitude as a function of redshift. The black dashed line is the SDSS $r=17.77$ magnitude limit and the blue line limits the $M_r<-20.5$ volume limited sample. (C) $R$-band $K$-correction for SHELS galaxies as a function of redshift. The blue line is the median $K$-correction in equally populated bins of redshift and the red line is a parabolic fit to the median correction. (D) Absolute $K$-corrected $R$-band magnitude as a function of redshift. The black dashed line is the SHELS $R = 20.6$ magnitude limit and the blue line limits the $M_R<-20.8$ volume limited sample.}
\label{fig:kcorr}
\end{figure*}

\subsection{Sample Selection}

We select SDSS galaxies with $M_\ast > 10^{9} M_\odot$, $0.02 < z < 0.2$ and $D_{n}4000 > 1.5$ to conservatively select quiescent galaxies dominated by older stellar populations \citep{Kauffmann2003a, Woods2010}. We do not explicitly limit the minimum $\sigma$ to avoid biasing the $M_\ast \sigma$ relation. However, our stellar mass limit ensures that only a small fraction of galaxies have velocity dispersions near the limit set by the SDSS spectrograph resolution. We remove a small fraction of galaxies ($<0.3\%$) with $\Delta >50$ km s$^{-1}$ and $\sim150$ outliers from the $M_\ast \sigma$ relation by visual inspection. The selection on $\Delta$ and rejection of outliers removes a very small fraction of the galaxies in the sample. This removal does not impact our results. The final sample consists of $\sim370,000$ galaxies. Figure \ref{fig:hist} shows the properties of the SDSS sample.

\begin{figure}
\begin{center}
\includegraphics[width = \columnwidth]{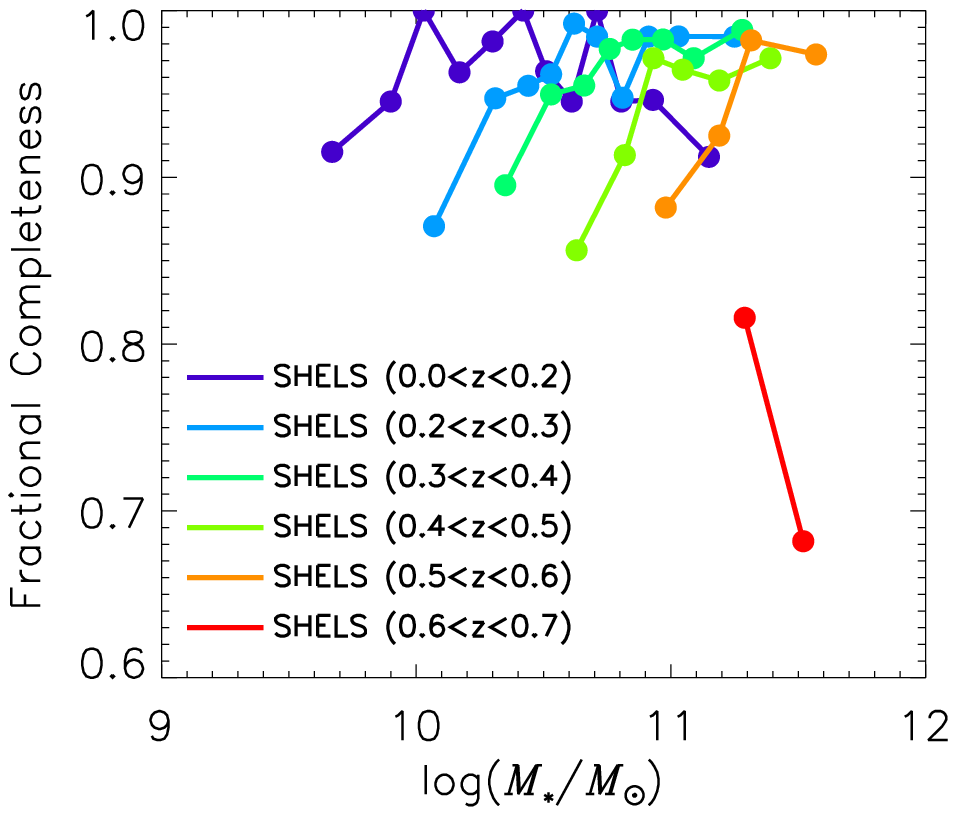}
\end{center}
\caption{The fraction of galaxies in the SHELS sample with velocity dispersion measurements as a function of stellar mass and redshift. The completeness is $\gtrsim90\%$ for most of the stellar mass and redshift range; at $0.6<z<0.7$ the data are incomplete due to the low surface brightness of the highest redshift objects.}
\label{fig:completeness}
\end{figure}

We limit the SHELS analysis to the ``complete" sample with $R<20.6$ \citep[see][]{Geller2014}. We select galaxies with $M_\ast>10^{9.5} M_\odot$ and $D_n4000 > 1.5$. The stellar mass limit ensures that only a small fraction of galaxies have velocity dispersions near the limit set by the Hectospec resolution. The final sample consists of 4585 galaxies. Figure \ref{fig:hist} shows the properties of the SHELS sample.

For the SHELS data, $\Delta$ is correlated with $\sigma$. To avoid biasing the $M_\ast \sigma$ relation, we make no S/N ratio or error cuts on the velocity dispersion measurement. Figure \ref{fig:completeness} shows the fraction of galaxies in the SHELS sample with velocity dispersion measurements. We measure a $\sigma$ for $95\%$ of the sample galaxies. At $0.6<z<0.7$ the incompleteness is substantial (see Figure \ref{fig:completeness}); uncontrolled systematic effects may bias the $M_\ast \sigma$ relation measured in this redshift range. Over the full redshift range, the SHELS ``complete" redshift sample is $\gtrsim90\%$ complete and thus we measure velocity dispersions for $>85\%$ of galaxies in the F2 field with $R<20.6$, $M_\ast>10^{9.5} M_\odot$ and $D_n4000 > 1.5$. The high completeness implies that any selection bias is mainly a direct result of the magnitude limited sample.

{We test the sensitivity of our results to the $D_n4000$ selection. Increasing the selection threshold to $D_n4000 > 1.6$ does not impact our results. We attribute this result to the fact that the $M_\ast \sigma$ relations in bins of $D_n4000$ are parallel (Zahid et al., In Preparation). Lowering the threshold $D_n4000 > 1.4$ results in lower mass bins having larger median velocity dispersions across all redshifts. We examine the spectra and images of galaxies contributing to this bias. The spectra all show emission lines and the images where the objects are well resolved appear to be disky. Thus, we attribute this larger median velocity dispersion in lower mass bins to the inclusion of star-forming, disk galaxies within the sample; for these galaxies the stellar kinematics are dominated by ordered rotation, not random motion. The star-forming galaxy fraction scales with stellar mass, therefore the effect is most prominent at lower stellar masses. We conclude that given the statistical uncertainties, the selection effects related to the $D_n4000$ cut we apply are small as long as we restrict our attention to the quiescent galaxy population.}

 \subsection{Volume Limited Sample}

We derive volume limited samples to test the robustness of our results. The SDSS and SHELS data are $r$-band and $R$-band selected, respectively; therefore we derive volume limited samples for the data sets independently. For the SDSS data we use the $r$-band $K$-corrections given in the Value Added Galaxy Catalog\footnote{http://cosmo.nyu.edu/blanton/vagc/kcorrect.html} of the NYU group \citep{Blanton2005a}. For the SHELS data, we derive the $R$-band $K$-correction based on the \citet{Blanton2007} code. To minimize the correction the SDSS and SHELS magnitudes are $K$-corrected to $z=0.1$ and $z=0.35$, respectively, adopting $H_0 = 70$ km s$^{-1}$.

Figure \ref{fig:kcorr}A and \ref{fig:kcorr}C show the $r$-band ($K_r$) and $R$-band ($K_R$) $K$-corrections, respectively, as a function of redshift. The best-fits to the median $K_r$ and $K_R$ as a function of redshift are
\begin{equation}
 K_r(z) = 1.26 \, z_{0.1}  + 1.52\,z^2_{0.1} - 2.5\, \mathrm{log_{10}}(1.1)
\label{eq:kcorr_sdss}
\end{equation}
and
\begin{equation}
 K_R(z) = 1.55 \, z_{0.35}  + 2.94\,z^2_{0.35} - 2.5\, \mathrm{log_{10}}(1.35),
\label{eq:kcorr_shels}
\end{equation}
respectively. Here $z_{0.1} = z - 0.1$ and $z_{0.35} = z - 0.35$. We derive the volume limited samples by $K$-correcting each galaxy and by applying the median corrections given in Equations \ref{eq:kcorr_sdss} and \ref{eq:kcorr_shels} to the magnitude limit of each survey. Figure \ref{fig:kcorr}B and \ref{fig:kcorr}D show the data and the volume limited selection limits of $M_r < -20.5$ and $M_R < -20.8$, respectively. Galaxies scatter across the magnitude limit due to photometric errors. The volume limited SDSS and SHELS samples consist of $\sim 110,000$ and 1827 galaxies, respectively.

\section{The Relation Between Stellar Mass and Velocity Dispersion}

\begin{figure}
\begin{center}
\includegraphics[width =  \columnwidth]{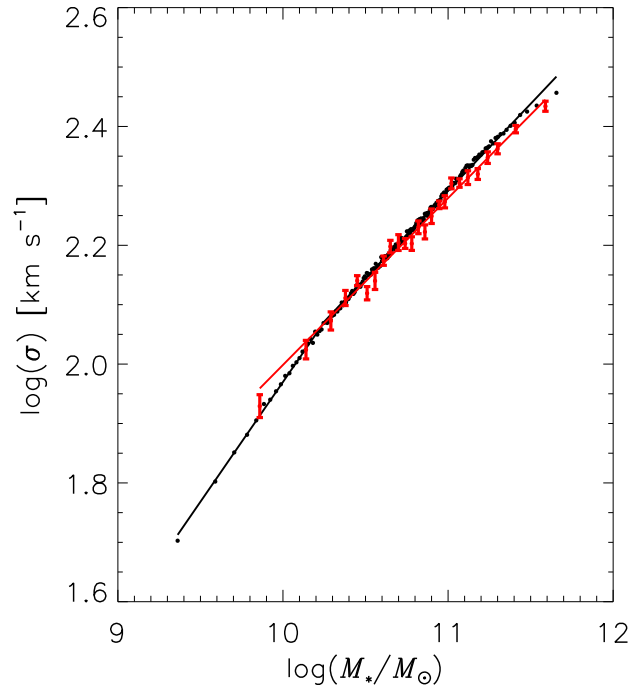}
\end{center}
\caption{Median velocity dispersion in bins of stellar mass for the SDSS (black points) and SHELS (red points) {magnitude limited} samples. The black and red lines show the best-fit to the SDSS and SHELS data, respectively (see Section 3). The errors are bootstrapped.}
\label{fig:fit}
\end{figure}

We derive the $M_\ast \sigma$ relation for the SDSS and SHELS samples independently. We calculate the median log$(\sigma)$ in equally populated stellar mass bins and bootstrap the errors. Figure \ref{fig:fit} shows the $M_\ast \sigma$ relation for the SDSS and SHELS samples.

Figure \ref{fig:fit} shows that the SDSS $M_\ast \sigma$ relation exhibits clear evidence of a break in the power law \citep[c.f.][]{Nigoche-Netro2011, Cappellari2013b, Cappellari2016}. We fit the data with
\begin{equation}
\sigma \left( M_\ast \right) = \sigma_b \left ( \frac{M_\ast}{M_b} \right)^{\alpha_1} \,\, \mathrm{for} \, \, \,M_\ast \leq M_b \nonumber
\end{equation}
\begin{equation}
\sigma \left( M_\ast \right) = \sigma_b \left ( \frac{M_\ast}{M_b} \right)^{\alpha_2} \,\, \mathrm{for} \, \, \, M_\ast > M_b.
\label{eq:fit}
\end{equation}
We parameterize the fit with a break point in stellar mass and velocity dispersion given by $M_b$ and $\sigma_b$ measured in units of $M_\odot$ and km s$^{-1}$, respectively. The two indices $\alpha_1$ and $\alpha_2$ define the power law below and above the break point, respectively. We fit the SDSS data in log-log space by minimizing $\chi^2$ using \emph{mpcurvefit.pro} implemented as part of the {\sc MPFIT IDL} package \citep{Markwardt2009}. Table \ref{tab:fit_param} lists the best-fit parameters and errors for the SDSS data. \citet{Cappellari2013b} report the characteristic break in the power law at a dynamical mass of $3\times10^{10} M_\odot$ {\citep[see also][]{Hyde2009a, Bernardi2011}}. Our measurement of $M_b$ is a factor of 1.6 smaller than \citet{Cappellari2013b}. However, we parameterize the relation in terms of stellar mass \emph{not} dynamical mass and thus we expect our measured value of the break point to be smaller due to the additional contribution from dark matter to the dynamical mass. {We note that \citet{Bernardi2011} report a second break from the single power law at log($M_\ast/M_\odot \sim 11.3$. We also observe this break at large stellar mass (e.g., see Figure \ref{fig:fit}).}


We fit the SHELS data with a single power law because the best-fit double power law yields a break point stellar mass outside the range of the data. The SHELS data do not preclude the double power law form but the statistical uncertainties and stellar mass range covered by the SHELS $M_\ast \sigma$ relation do not require such a parameterization to describe the data.  We fix $M_b$ and fit for $\sigma_b$ and $\alpha_2$ using the \emph{linfit.pro} routine in {\sc IDL}. Table \ref{tab:fit_param} gives the best-fit parameters and errors. We can directly compare the fits to the SHELS and SDSS data. The zero points are consistent. The slopes differ by $\lesssim2.4$ standard deviations but the difference is small enough that it is probably solely dominated by residual systematics.

\begin{figure}
\begin{center}
\includegraphics[width =  \columnwidth]{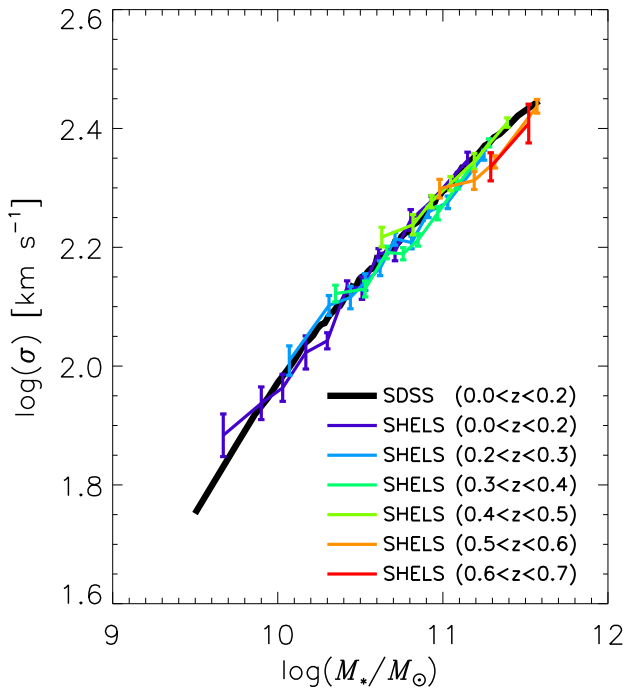}
\end{center}
\caption{Median relation between stellar mass and velocity dispersion in bins of redshift {for the magnitude limited samples}. The errors are bootstrapped. Note the absence of evolution with redshift.}
\label{fig:vdisp_evol}
\end{figure}

We examine the redshift evolution of the $M_\ast \sigma$ relation by sorting the SHELS data into bins of redshift. Figure \ref{fig:vdisp_evol} shows the $M_\ast \sigma$ relation for {$z<0.7$}. Over the stellar mass and redshift ranges probed by the SHELS sample, the $M_\ast \sigma$ relation does not appear to depend on redshift.

\begin{figure*}
\begin{center}
\includegraphics[width = 1.6\columnwidth]{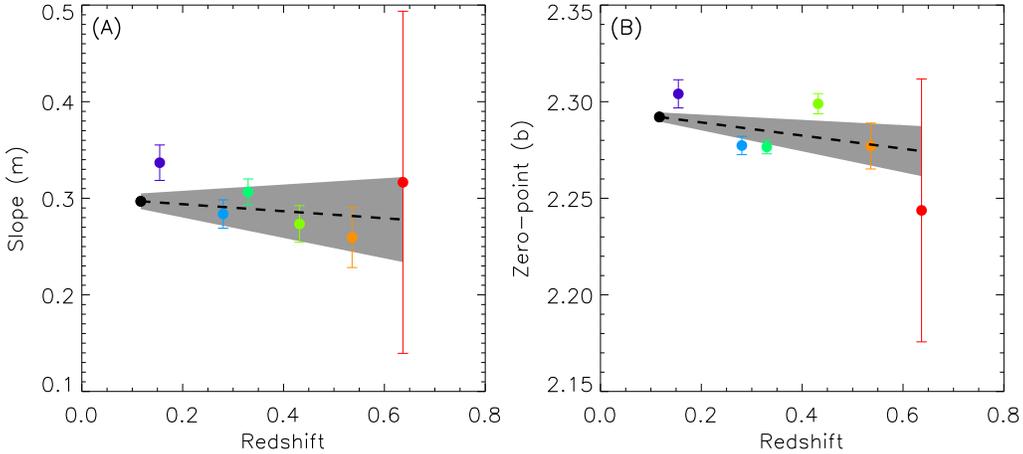}
\end{center}
\caption{Best-fit power law (A) slope and (B) and zero point of the $M_\ast \sigma$ relation as a function of redshift {for the magnitude limited samples}. The gray region indicates the $2\sigma$ uncertainty of the fit. The fit parameters are consistent with no redshift evolution.}
\label{fig:fitz}
\end{figure*}

We quantify the redshift evolution by fitting the $M_\ast \sigma$ relation in bins of redshift. We limit the SDSS sample to galaxies with $M_\ast > M_b$ where log$(M_b/M_\odot) = 10.26$ is the break point stellar mass of the double power law fit to the SDSS data. By limiting the SDSS mass range, we can fit all the $M_\ast \sigma$ relations with the single power law model. Figure \ref{fig:fitz}A and \ref{fig:fitz}B show the slope and zero point as a function of redshift. The parameters are derived from $\chi^2$-fitting the $M_\ast \sigma$ relations shown in Figure \ref{fig:vdisp_evol}. The slope, $m$, and zero point, $b$, as a function of redshift are
\begin{equation}
m(z) = 0.301 \pm 0.004 - (0.036 \pm 0.035) \, z
\end{equation}
and
\begin{equation}
b(z) =  2.296 \pm 0.001 - (0.034 \pm 0.010) \,z,
\end{equation}
respectively. {The redshift dependence of the $M_\ast \sigma$ slope is consistent with zero. Over the redshift range of our data, the zero-point evolves by $\sim0.02$ dex. This small amount of evolution is at a level that is likely dominated by systematic uncertainties. If we weight each data point in the fit equally rather than by the observational uncertainty, no statistically significant evolution appears. The $M_\ast \sigma$ relation for massive galaxies does not depend strongly on redshift for galaxies at $z<0.7$.}

\begin{figure*}
\begin{center}
\includegraphics[width = 1.6\columnwidth]{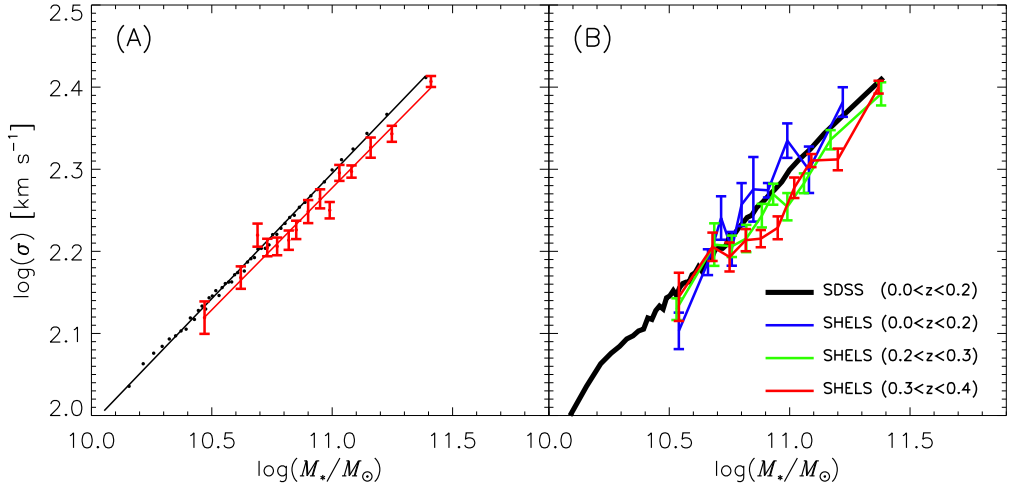}
\end{center}
\caption{(A) The $M_\ast \sigma$ relation for the SDSS (black points) and SHELS (red points) derived from the volume limited samples. The black and red lines are the best-fit relation. (B) The $M_\ast \sigma$ relation as a function of redshift derived from the volume limited samples. The error bars are bootstrapped.}
\label{fig:fit_rlim}
\end{figure*}

To check for systematics in the $M_\ast \sigma$ relation introduced by the magnitude limit we examine the volume limited samples. As before, we calculate the median velocity dispersion in bins of stellar mass and bootstrap the errors. Figure \ref{fig:fit_rlim}A shows the $M_\ast \sigma$ relation. The SDSS and SHELS $M_\ast \sigma$ relations are consistent within the errors and both are consistent with the relations derived from the full sample. {The offset in Figure \ref{fig:fit_rlim}A is $\sim0.01$ dex, an effect we attribute to residual systematics.} Figure \ref{fig:fit_rlim}B shows the $M_\ast \sigma$ relations derived from the volume limited sample as a function of redshift. The $M_\ast \sigma$ remains independent of redshift.


 \section{Scatter in the Relation Between Stellar Mass and Velocity Dispersion}
 
\begin{figure*}
\begin{center}
\includegraphics[width = 2 \columnwidth]{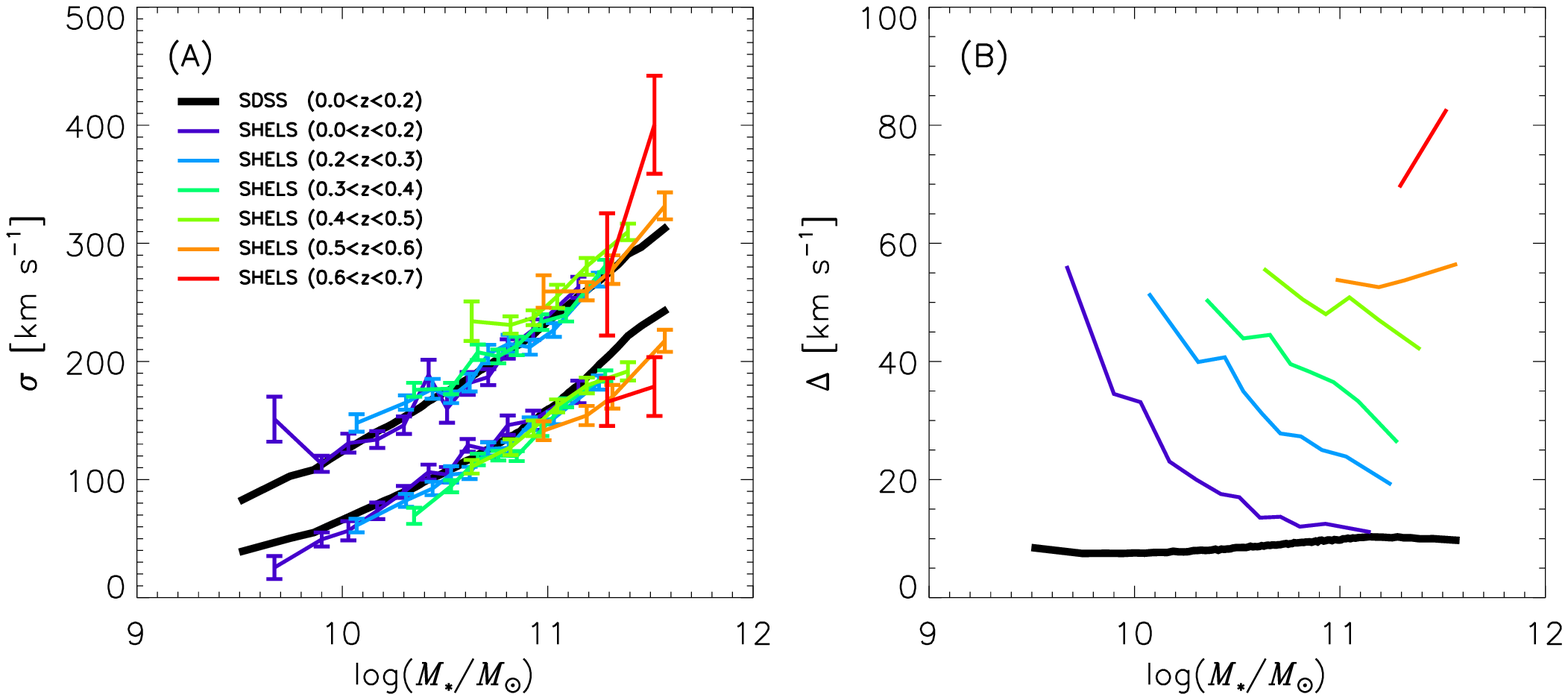}
\end{center}
\caption{(A) {Limits} of the central 68\% of the velocity dispersion distribution in bins of redshift and stellar mass derived from the SDSS and SHELS samples. The errors are bootstrapped. (B) Median error in the velocity dispersion for individual galaxies as a function of stellar mass and redshift.}
\label{fig:scatter}
\end{figure*}
 
Figure \ref{fig:scatter}A shows the central 68\% of the velocity dispersion distribution as a function of stellar mass and redshift. As before, we bin the SHELS data by stellar mass and redshift and determine errors by bootstrapping the data. The data show a weak trend of increasing width in the velocity dispersion distribution at a fixed stellar mass with redshift. We largely attribute this trend to observational uncertainties. 

Figure \ref{fig:scatter}B shows the median $\Delta$ in bins of stellar mass and redshift. The median $\Delta$ tends to decrease with stellar mass at fixed redshift. At a fixed stellar mass, galaxies at greater redshifts tend to be less bright, thus $\Delta$ increases with redshift. \citet{Zahid2016} show that larger $\Delta$s artificially broaden the width of the $\sigma$ distribution as expected. We attribute the larger width of the $\sigma$ distribution at greater redshifts to the increasing $\Delta$s.

\begin{figure*}
\begin{center}
\includegraphics[width = 2 \columnwidth]{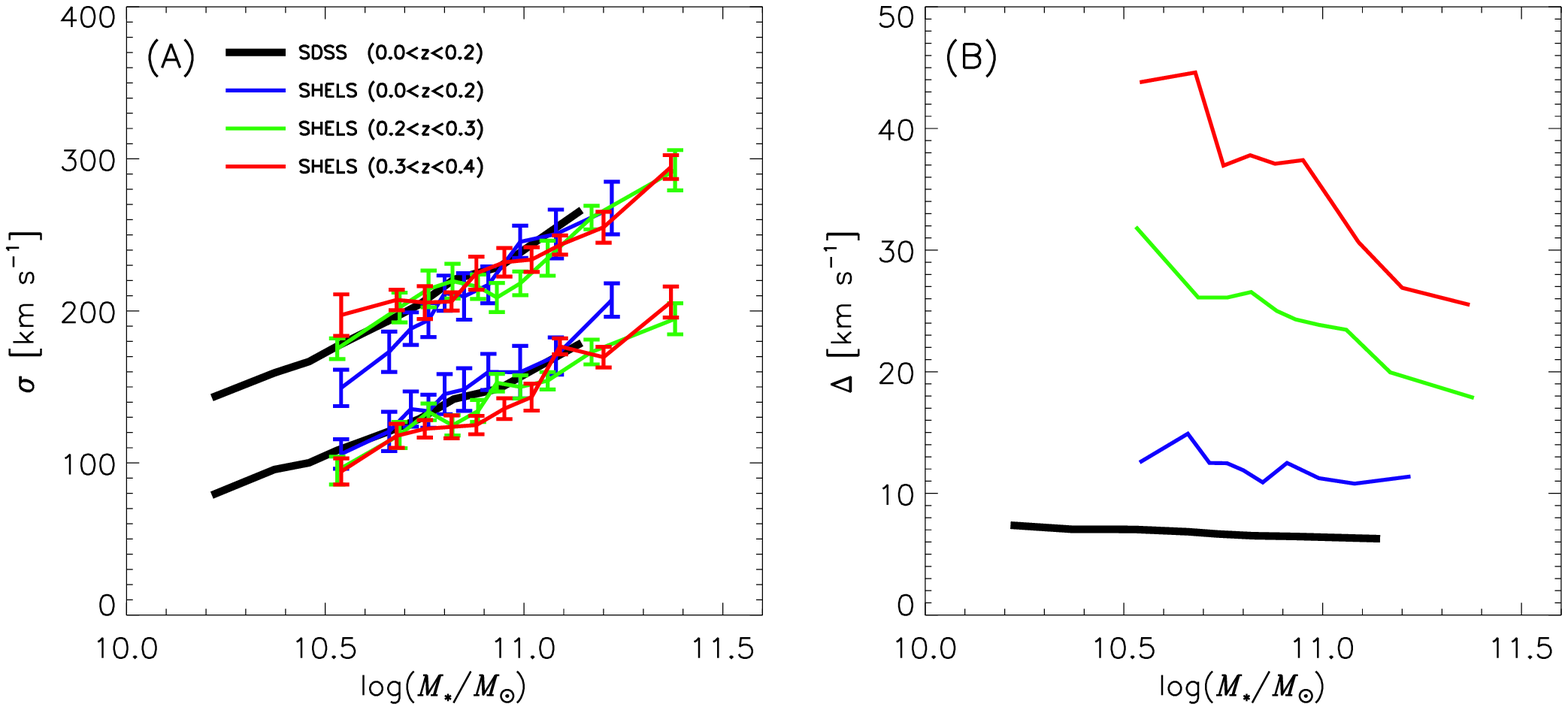}
\end{center}
\caption{(A) {Limits} of the central 68\% of the velocity dispersion distribution in bins of redshift and stellar mass derived from the SDSS and SHELS volume limited samples. The errors are bootstrapped. (B) Median error in the velocity dispersion for individual galaxies as a function of stellar mass and redshift.}
\label{fig:scatter_rlim}
\end{figure*}

We test our results by examining the width of velocity dispersion distribution of the volume limited samples. Figure \ref{fig:scatter_rlim}A shows the limits delineating the central 68\% of the velocity dispersion and Figure \ref{fig:scatter_rlim}B shows the median $\Delta$ in bins of stellar mass and redshift. The width of the $\sigma$ distribution in the volume limited at $z<0.4$ appears to be independent of redshift. The $\Delta$s for the volume limited sample are significantly smaller than the full SHELS sample because of the limited stellar mass and redshift range. This result for the volume limited samples indicates further that the apparent evolution of the width of the velocity dispersion distribution results from larger $\Delta$s for galaxies at lower stellar masses and/or higher redshifts in the full magnitude limited sample. We note that \citet{Shu2012} report a comparably small variation in the intrinsic width of the velocity dispersion distribution for massive galaxies at $z<0.7$.

\section{Systematic Issues}

There are several systematic issues that may affect the relation between stellar mass and velocity dispersion. They are i) use of $\sigma$ or galaxy properties derived from $\sigma$ as the independent variable and/or ii) selection of the data based on $\Delta$. Both of these approaches may introduce spurious trends with redshift which appear as evolution when there is none.

\subsection{Choice of Independent Variable}

{Luminosity (or stellar mass) has long been recognized as the important independent variable when examining velocity dispersions \citep[e.g.,][]{Schechter1980}. Here we highlight and discuss the consequences of taking velocity dispersion as the independent variable in the context of magnitude limited surveys.}

We measure the median $\sigma$ as a function of stellar mass. This approach is critical because the SHELS survey is magnitude limited and therefore at a fixed stellar mass, the velocity dispersion distribution is marginally biased for a sample of quiescent galaxies. In contrast, for a magnitude limited survey the stellar mass distribution may be biased at a fixed $\sigma$.

\begin{figure*}
\begin{center}
\includegraphics[width = 2 \columnwidth]{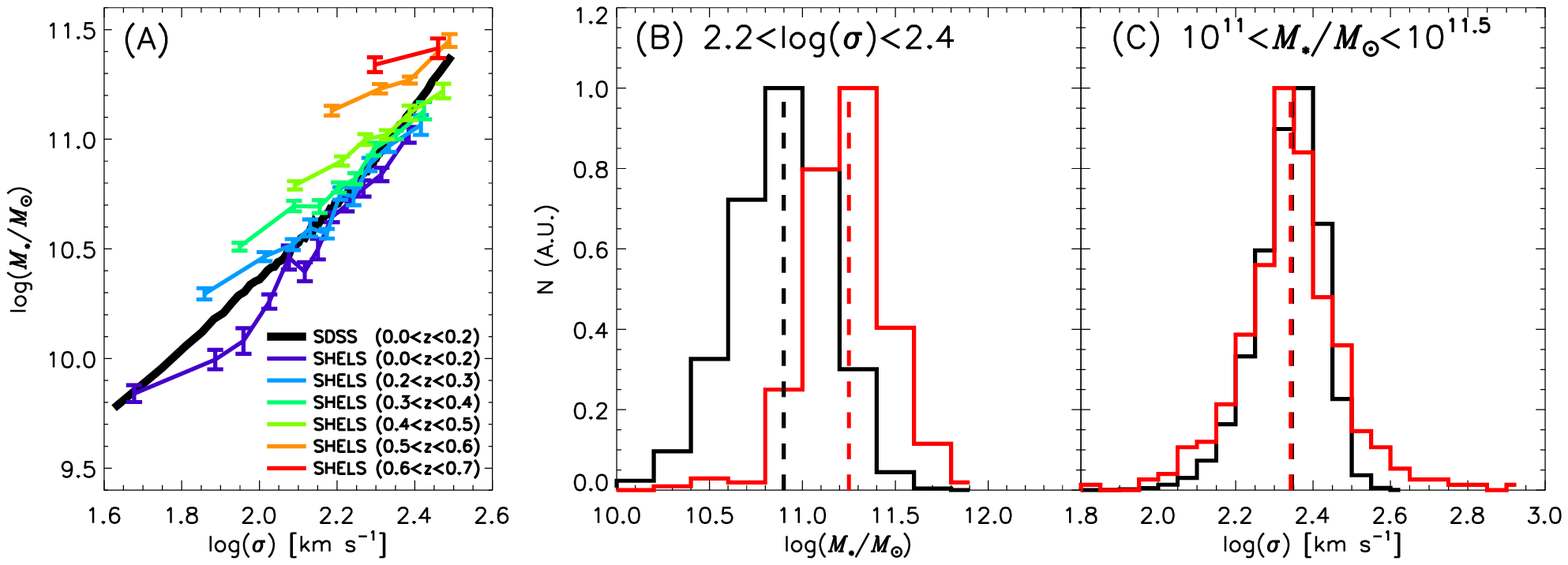}
\end{center}
\caption{(A) Median stellar mass in bins of velocity dispersion and redshift for the SDSS and SHELS samples. Here velocity dispersion is the independent variable. Note the apparent evolution in the relation between velocity dispersion and stellar mass. (B) Stellar mass distribution for the SDSS (black) and SHELS (red) samples in a narrow bin of velocity dispersion. The SHELS sample is limited to $z>0.5$. The black and red dashed lines are the median of the SDSS and SHELS stellar mass distributions, respectively. Note the separation in the two distributions. (C) Velocity dispersion distribution for the SDSS (black) and SHELS (red) samples in a narrow bin of stellar mass. The SHELS sample is limited to $z>0.5$. Black and red dashed lines are the median of the SDSS and SHELS velocity dispersion distributions, respectively. The two medians are nearly coincident.}
\label{fig:sigma_mass}
\end{figure*}

Figure \ref{fig:sigma_mass}A shows the median stellar mass as a function of $\sigma$. Taken at face value, the $\sigma M_\ast$ relation shows strong redshift evolution. This apparent evolution is a consequence of selection. Figure \ref{fig:sigma_mass}B shows the stellar mass distribution in a narrow $\sigma$ range for the SDSS and for SHELS galaxies at $z>0.5$. The median of the stellar mass distribution of SHELS galaxies is offset to significantly larger stellar masses. Conversely, Figure \ref{fig:sigma_mass}C shows the velocity dispersion distribution in a narrow stellar mass range for SDSS and for SHELS galaxies at $z>0.5$. The medians of the two distributions are consistent with one another. The SHELS distribution is broader as a consequence of the larger $\Delta$s \citep[c.f.][]{Zahid2016}.

SHELS is a magnitude limited survey. At a fixed $\sigma$, the magnitude range$-$and consequently the stellar mass distribution$-$is not fully sampled. The mass-to-light ratio for galaxies with $D_n4000>1.5$ does not vary significantly from one galaxy to another \citep[see Figure 12 in][]{Geller2014} and thus in narrow bins of redshift the observed magnitude is nearly a direct proxy for the stellar mass. The velocity dispersion distribution is essentially complete above this stellar mass threshold because there is no a priori selection on $\sigma$. However, the stellar mass distribution is \emph{not} complete in a narrow range of $\sigma$ because the survey is magnitude limited. Figure \ref{fig:sigma_mass}C emphasizes the importance of using the stellar mass or luminosity as the \emph{independent} variable when examining  evolution of relations between stellar mass and velocity dispersion, or quantities derived from the velocity dispersion.

\subsection{Selection Based on Data Quality}

\begin{figure*}
\begin{center}
\includegraphics[width = 2 \columnwidth]{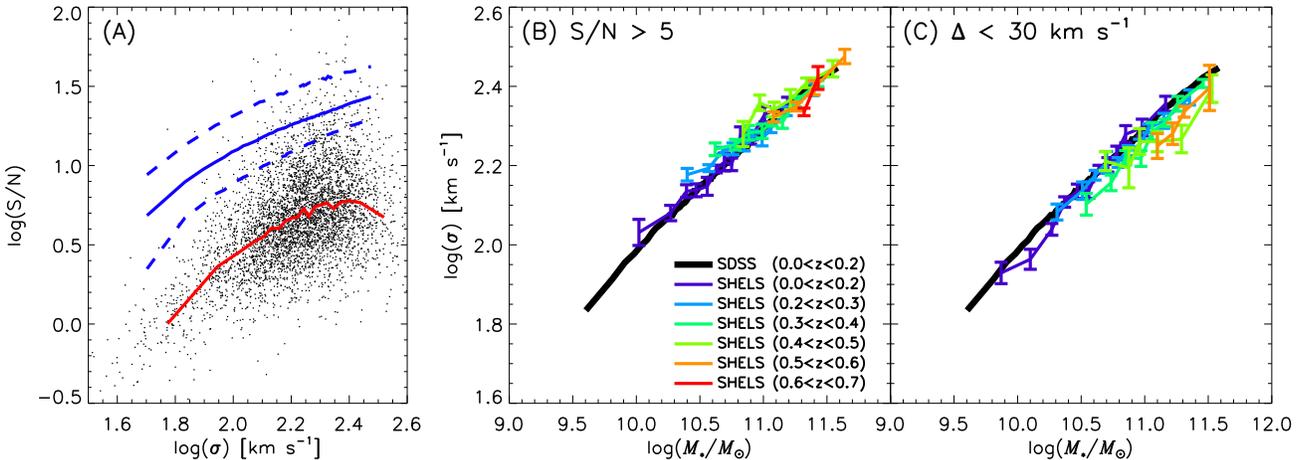}
\end{center}
\caption{(A) {S/N ratio as a function of velocity dispersion for the SHELS and SDSS samples. The black points are individual SHELS galaxies and the solid red line is the median S/N ratio of SHELS galaxies in bins of velocity dispersion. The blue solid and dashed lines are the median S/N ratio and the limits containing 80\% of the S/N distribution of SDSS galaxies in bins of velocity dispersion, respectively.} (B) The $M_\ast \sigma$ relation in bins of redshift for measurements with S/N ratio $> 5$. (C) The $M_\ast \sigma$ relation in bins of redshift for galaxies with $\Delta<30$ km s$^{-1}$.}
\label{fig:sncut}
\end{figure*}

Treatment of errors in individual measurements may also introduce spurious trends in the $M_\ast \sigma$ relation. Figure \ref{fig:sncut}A shows that the S/N ratio in an individual measurement of velocity dispersion is correlated with the value of the velocity dispersion {for both the SHELS and SDSS samples.} Galaxies with larger $\sigma$s tend to be brighter at a fixed redshift and therefore the corresponding $\Delta$s are typically smaller. 

Figure \ref{fig:sncut}B shows the $M_\ast \sigma$ relation derived by imposing a S/N ratio cut on the SHELS sample. Because of the correlation between velocity dispersion and S/N ratio, a S/N ratio cut biases the sample towards larger velocity dispersion objects. The bias, however, is not severe. It affects the lower stellar mass bins at each redshift interval {and is roughly analogous to imposing a new magnitude limit.} 

Figure \ref{fig:sncut}C shows the $M_\ast \sigma$ relation derived by selecting galaxies with $\Delta < 30$ km s$^{-1}$. For the SHELS data, imposing a maximum cut on $\Delta$ leads to a spurious trend which appears as redshift evolution, {albeit the trend is weak. At a fixed stellar mass, $\Delta$ is correlated with $\sigma$, i.e. galaxies with larger velocity dispersions typically have larger absolute errors. Therefore, a hard cut on $\Delta$ leads to preferential removal of high $\sigma$ galaxies. Because $\Delta$ is also correlated with redshift, this effect is more prominent for the higher redshift bins.}

The impact that measurement quality selection has on the derived relations depends on the survey observing strategy {and data quality.  SDSS galaxies have velocity dispersions measured at S/N ratios that are typically $0.6\sim0.7$ dex higher than the SHELS data. Thus, the S/N ratio $>5$ and $\Delta < 30$ km s$^{-1}$ cuts do not impact the SDSS sample as significantly as the SHELS sample.}

We do not apply any selection based on $\Delta$. Figure \ref{fig:sncut} shows that the redshift independence of the $M_\ast \sigma$ relation for massive galaxies at $z<0.7$ is reasonably robust to a S/N ratio cut but {but is susceptible to bias when hard cuts on $\Delta$ are applied. The impact of these types of cuts depends on the quality of the data.}


\section{Discussion}

Scaling relations provide important constraints for understanding galaxy formation and evolution. {We measure the local relation between stellar mass and velocity dispersion from SDSS; this relation is consistent with the previous measurement from \citet[see Table \ref{tab:fit_param}]{Hyde2009a}. Using the SHELS sample, we show that this relation is independent of redshift for massive galaxies at $z<0.7$. These measurements provide constraints for understanding galaxy evolution.} 

\citet{Cappellari2016} provides a detailed review of galaxy scaling relations measured from integral field spectroscopy (IFS) which allow for stringent constraints on the physical processes of early-type galaxy formation. The central stellar velocity dispersion measurements we use are not spatially resolved and thus cannot provide the exquisite detail made possible by the IFS. However, IFS observations are only available for small samples of nearby galaxies whereas central velocity dispersions can be measured for large samples out to high redshifts. Here we focus our discussion on the potential applications of large samples of central velocity dispersions. We refer readers to \citet[and many references therein]{Cappellari2016} for a comprehensive discussion of the physical processes governing the structural and kinematic properties of early-type galaxies measured using IFS.

The galaxy observable that best connects galaxies with their dark matter halos has important implications for the study of galaxies and cosmology. From an observational perspective, velocity dispersion is a significantly more expensive measurement than a stellar mass derived from photometry. However, stellar masses suffer from systematic issues and model dependencies \citep{Conroy2009a, Behroozi2013b} whereas velocity dispersions are a directly measured property of galaxies. Velocity dispersions do require careful calibration of the measurement technique \citep[e.g.,][]{Fabricant2013}. 

The $M_\ast \sigma$ relation for massive galaxies at $z<0.7$ is independent of redshift. Thus, stellar mass and velocity dispersion provide equally valid descriptions of galaxies \emph{on average}. {However, in detail, the existence of a fundamental plane \citep[][]{Djorgovski1987, Dressler1987} means the two are not one-to-one.} A simple interpretation of these results is that over the stellar mass and redshift range examined here, the majority of the quiescent galaxy population is already well in place and has not evolved significantly. This conclusion is consistent with the little evolution of the massive quiescent galaxy stellar mass function at $z\lesssim0.7$ \citep{Moustakas2013, Ilbert2013, Muzzin2013b}. 

Several studies have examined whether stellar mass or velocity dispersion is a more fundamental property of galaxies \citep[e.g.,][]{Bernardi2005, Wake2012a, Li2013, vanUitert2013}. No consensus has been established partly because there has been no comprehensive theoretical study. Such a study based on hydrodynamic simulations could in principle forge a theoretical link between the observed central stellar velocity dispersion and the velocity dispersion of the dark matter halo. It is critical that the simulations attempt to match observations as well as possible. For example, stellar velocity dispersions should be the projected line-of-sight velocity dispersions determined in a cylinder. 

\subsection{The Redshift Evolution and Scatter Around the $M_\ast \sigma$ Relation}

The scatter relative to the median $M_\ast \sigma$ relation is correlated with other galaxy properties including, e.g., galaxy size and/or age. There is, for example, a well known relation between stellar mass density, velocity dispersion and size$-$the stellar mass fundamental plane \citep[e.g.,][]{Hyde2009b}. This plane does not appear to evolve significantly for galaxies at $z<0.6$ \citep{Zahid2016}. The scatter around the $M_\ast \sigma$ relation also appears to be correlated with galaxy age \citep{Forbes1998, Bernardi2005}.

\begin{figure}
\begin{center}
\includegraphics[width = \columnwidth]{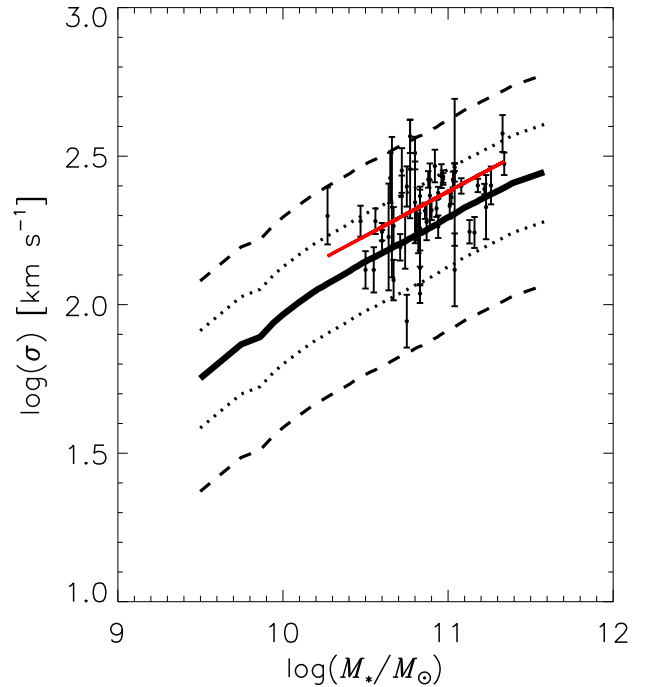}
\end{center}
\caption{Velocity dispersions of 56 galaxies (black points and error bars) as a function of stellar mass at $0.9<z<1.6$ from \citet{Belli2014}. The solid line is the median SDSS velocity dispersion in bins of stellar mass. The dotted and dashed lines are the 68 and 95\% contours of the SDSS velocity dispersion distribution as a function of stellar mass. The red curve is the best-fit $M_\ast \sigma$ relation for the \citet{Belli2014} sample. Note the apparent evolution in the $M_\ast \sigma$ relation at $z>1$.}
\label{fig:belli}
\end{figure}

Observations of high redshift galaxies suggest the $M_\ast \sigma$ relation may evolve at $z>0.7$. Figure \ref{fig:belli} shows the $M_\ast \sigma$ relation at $0.9<z<1.6$ from \citet{Belli2014}. The slope of the relation is consistent with our measurements at $z<0.7$; the zero point is significantly offset (see Table \ref{tab:fit_param}). Taken at face value, these results suggest significant evolution of the $M_\ast \sigma$ relation at higher redshift. \citet{Belli2014} show that the relation between stellar and dynamical mass for these data follow the local relation more closely and therefore attribute the offset to the smaller sizes of galaxies at higher redshifts. This result underscores the fact that the scatter in the $M_\ast \sigma$ relation is correlated with other galaxy properties including their size. When examined in detail, it is clear that stellar mass and velocity dispersion do not provide equivalent descriptions of galaxies; one of these properties is likely to be more closely connected to the dark matter halo.

\subsection{Connecting Galaxies with their Dark Matter Halos}

A standard statistical approach for connecting observable galaxies to their dark matter halos is the abundance matching technique \citep{Vale2004}. In its simplest form, the technique assigns galaxies observed within a limited volume to dark matter halos in a volume-matched N-body simulation assuming that rank order is preserved, i.e. the most luminous (or massive) galaxy observed is associated with the most massive simulated halo, the second most luminous galaxy is associated with the second most massive halo and so on. Abundance matching reproduces many statistical properties of the galaxy population \citep[e.g.,][]{Conroy2006, Guo2010, Behroozi2013a, Hwang2016}; however, in detail, the assumption that rank order is preserved may be violated \citep{Leja2013, Mundy2015, Torrey2015}. Velocity dispersions may provide a better means to connect galaxies with dark matter halos. This possibility highlights the need for derivation of the central dark matter and stellar velocity dispersions from hydrodynamical simulations in a manner consistent with observations, i.e. the projected line-of-sight velocity dispersions determined in circular apertures.

Remarkably the power law scaling of the stellar mass and velocity dispersion is consistent with the expected scaling between the dark matter velocity dispersion and the total halo mass \citep{Evrard2008, Posti2014}. \citeauthor{Evrard2008} show that $\sigma_{DM} \propto M^{0.33}_{200}$ over a range of halo masses that includes both individual galaxies and galaxy clusters (see their Figure 6). Here $\sigma_{DM}$ is the dark matter velocity dispersion and $M_{200}$ is the total mass enclosed within the virial radius. The consistency between the slope of the $M_\ast \sigma$ relation we derive and the slope expected for virialized dark matter halos is surprising. The central stellar velocity dispersion is a tracer of the total central dynamical mass including centrally concentrated dark matter. However, the relation we derive is between \emph{stellar mass} and velocity dispersion. If the galaxies we examine are virialized systems, the scaling we derive suggests that either there is a negligible central dark matter contribution or the central stellar-to-dark matter ratio does not depend strongly on stellar mass. In other words, the observed central stellar velocity dispersion is simply proportional to the velocity dispersion of the dark matter halo.

\begin{figure}
\begin{center}
\includegraphics[width =  \columnwidth]{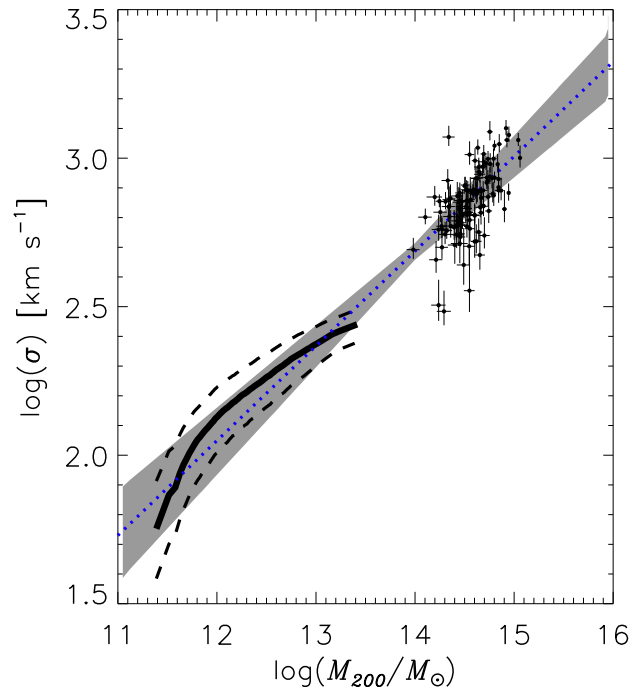}
\end{center}
\caption{Velocity dispersion as a function of the total mass, $M_{200}$. The black points are the Sunyaev-Zeldovich-selected (SZs) clusters from \citet{Rines2016}. The cluster velocity dispersion is determined from redshifts of cluster members and the total masses are the Sunyaev-Zeldovich masses are derived from \emph{Planck} satellite observations. The dotted blue line is the best-fit power law from \citet{Rines2016} and the gray region denotes the 68\% confidence interval. The solid and dashed black lines are the median and 68\% contours of the SDSS velocity dispersion distribution as a function of $M_{200}$ where $M_{200}$ for galaxies is the sum of the stellar and dark matter halo mass based on the stellar-mass-to-halo-mass relation from \citet{Behroozi2013a}. {The stellar-to-total mass fraction is consistent with a recent determination from weak lensing \citep{Bahcall2014}.} Note the remarkable overlap between the extrapolated cluster relation and the galaxy locus.}
\label{fig:rines}
\end{figure}

Following \citet{Evrard2008}, we compare the result for individual galaxies with the galaxy cluster scaling relations. \citet{Rines2016} examine the relation between total cluster mass determined using the Sunyaev-Zeldovich (SZ) effect \citep{Sunyaev1972} and the projected velocity dispersion measured from cluster members. Their measured relation is consistent with the theoretical relation from \citet{Evrard2008}. The slope we measure for the $M_\ast \sigma$ relation is consistent with the \citeauthor{Rines2016} measurement for their cluster sample; the zero point for the \citeauthor{Rines2016} sample is, however, significantly larger (see Table \ref{tab:fit_param}). The larger zero point is expected because our stellar mass estimate does not account for the dark matter. To compare our results with \citeauthor{Rines2016}, we calculate the total mass of galaxies in the SDSS sample using the \citet[private communication P.B.]{Behroozi2013a} stellar-to-halo mass conversion. 

Figure \ref{fig:rines} shows the central LOS stellar velocity dispersion for SDSS galaxies plotted against their total mass (stellar + dark matter) along with the SZ cluster results of \citet{Rines2016}. The locus of the SDSS galaxies falls firmly on the extrapolated best-fit relation (dotted blue line) of \citet{Rines2016}. This consistency has important implications for understanding the central stellar velocity dispersions of individual galaxies. The cluster velocity dispersion derived by \citeauthor{Rines2016} is sensitive to the total dynamical mass of the system. The fact that the relation between central stellar velocity dispersion and total mass for galaxies is consistent with \citeauthor{Rines2016} relation suggests that the central stellar velocity dispersion may be an unbiased tracer of the total mass of the galaxy. This conclusion is consistent with the strong-lensing results of \citet{Bolton2008}. They find that the ratio between the stellar velocity dispersion and the velocity dispersion inferred from lensing assuming an isothermal halo model is near unity. A one-to-one correspondence between the central stellar velocity dispersion and the central dark matter halo velocity dispersion would have very deep implications for connecting observations and theory.

\subsection{Variations in the Initial Mass Function}

We explicitly assume a universal IMF. Modifications of this assumption would change the results. Several recent studies report a non-universal IMF with variations that scale with the central stellar velocity dispersion in individual galaxies \citep{Treu2010, vanDokkum2010b, Cappellari2012, LaBarbera2013}; this reported variation in the IMF remains tentative \citep{Smith2014, Smith2015, Peacock2014}. If the IMF does vary systematically with velocity dispersion, the slope and zero point of the $M_\ast \sigma$ relation would change. In this case, the fact that the $M_\ast \sigma$ relation we measure is consistent with the virial scaling relation would be a mere coincidence. The SHELS data provide no constraint on the IMF variations. However, barring a conspiracy, the lack of evolution observed in the $M_\ast  \sigma$ relation does rule out any strong redshift evolution in the IMF at fixed velocity dispersion for massive galaxies at $z<0.7$.

\section{Summary and Conclusions}

We examine the relation between stellar mass and central stellar velocity dispersion$-$the $M_\ast \sigma$ relation$-$for massive quiescent galaxies at $z<0.7$ using data from SDSS and SHELS \citep[cf.][]{Montero-Dorta2016}. We cross-calibrate measurements of stellar mass and velocity dispersion for these two surveys. The $M_\ast \sigma$ relation and its scatter are both independent of redshift. 

The central stellar velocity dispersion and stellar mass may provide comparable descriptions of galaxies at $z<0.7$. However, the scatter in the relation between stellar mass and velocity dispersion correlates with other galaxy physical properties. There is also tentative evidence that the zero point of the relation evolves at $z>1$. In detail, the velocity dispersion and stellar mass are not equivalent descriptors. Determining which of these properties is more closely related to the dark matter halo has important implications for understanding galaxy evolution and cosmology.

The scaling between stellar mass and velocity dispersion may provide important clues for connecting early-type galaxies to a broader class of objects. The $M_\ast \sigma$ relation goes as $\sigma \propto M_\ast^{0.3}$. This scaling is expected for virialized dark matter halos, i.e. $\sigma_{DM} \propto M_{200}^{0.33}$ where $\sigma_{DM}$ is the dark matter halo velocity dispersion and $M_{200}$ is the total virial mass \citep{Evrard2008}. This similarity is somewhat surprising because there is no a priori expectation that the stellar mass and the stellar velocity dispersion should scale in the same manner as the dark matter halo mass and the dark matter halo velocity dispersion. We infer the dark matter halo mass of galaxies using a standard stellar-to-halo-mass conversion \citep{Behroozi2013a}. The relation between the total galaxy mass (stellar + dark matter) and the stellar velocity dispersion is surprisingly consistent with an extrapolation of the relation between the total mass of a galaxy cluster and its velocity dispersion \citep{Rines2016}. This result suggests that the stellar velocity dispersion may be directly proportional to the dark matter halo velocity dispersion.

Our results set the stage for more detailed explorations of the connection between stellar velocity dispersions and the properties of dark matter halos. The velocity dispersion may be a more fundamental observable than stellar mass and may provide a more robust means for connecting dark matter halos in N-body simulations with observed galaxies. A more direct connection between observations and theory could be based on a targeted analysis of large cosmological hydrodynamic simulations like the Illustris, EAGLE and MassiveBlack-II projects \citep{Vogelsberger2014a, Schaye2015, Khandai2015}. The luminosity-weighted two dimensional line-of-sight central velocity dispersion of stellar particles within a circular aperture (i.e. the projected velocity dispersion measured within a cylinder in analogy with the observations) could be compared with the appropriately measured velocity dispersion of particles characterizing the dark matter halo. The relationship between these two velocity dispersions derived from the simulations could provide a guide for better interpretation of the observations.

\begin{deluxetable*}{llcccc}
\tablewidth{500pt}
\tablecaption{Fit Parameters}
\tablehead{\colhead{Sample} & \colhead{N} &\colhead{log($M_b/M_\odot$)} &\colhead{log($\sigma_b$) [km s$^{-1}$]} &\colhead{$\alpha_1$} & \colhead{$\alpha_2$} }
\startdata

\cutinhead{The full sample (Figure \ref{fig:fit}) }

SDSS 	& 371884		     &  $10.26 \pm 0.01$ & $2.073 \pm 0.003$ &    $  0.403 \pm 0.004 $&     $0.293 \pm 0.001$ \\
SHELS      &     4585                &   10.26                   & $2.071 \pm 0.004$ &                                  &     $0.281\pm0.005$    \\

\cutinhead{The full sample as a function of redshift (Figure \ref{fig:fitz})}

SDSS ($M_\ast > 10^{10.26} M_\odot$) &325186 & 11 &   $2.2969 \pm 0.0006$   && $0.299 \pm 0.001$\\  
SHELS ($0<z<0.2$)       & 412           & 11 &    $2.304 \pm 0.007$      && $0.34 \pm 0.02$\\
SHELS ($0.2<z<0.3$)    & 1115           & 11 &     $2.277 \pm 0.005$ 	&& $0.28 \pm 0.01$\\
SHELS ($0.3<z<0.4$)     & 1316          & 11 &    $2.277 \pm  0.004$ 	&& $0.31\pm 0.01$\\
SHELS ($0.4<z<0.5$)     & 818          & 11 &    $2.299 \pm 0.005$ 	&& $0.27\pm0.02$\\
SHELS ($0.5<z<0.6$)    & 443 		& 11 &     $2.28 \pm 0.01$   	&& $0.26 \pm 0.03$\\
SHELS ($0.6<z<0.7$)	& 61	& 11 &     $2.24 \pm 0.07$  	&& $0.32 \pm 0.18$\\

\cutinhead{The volume limited sample (Figure \ref{fig:fit_rlim})}\\

SDSS VL       &   109330               & 10.26                & $2.068 \pm 0.001$ &                                    &    $0.305 \pm 0.001$ \\
SHELS VL     & 1827                &  10.26                & $2.06 \pm 0.01$     &                                    &    $0.30 \pm 0.01$  \\
SHELS VL ($0<z<0.2$)    & 233         & 10.26 &    $2.03 \pm 0.02$      && $0.37 \pm 0.03$\\
SHELS VL ($0.2<z<0.3$)  & 641        & 10.26 &    $2.05 \pm 0.01$ 	&& $0.30 \pm 0.02$\\
SHELS VL ($0.3<z<0.4 $)   & 953       & 10.26 &    $2.03 \pm  0.01$ 	&& $0.33\pm 0.02$\\

\cutinhead{The \citeauthor{Hyde2009a}, \citeauthor{Belli2014} and \citeauthor{Rines2016} samples (Figures \ref{fig:belli} and \ref{fig:rines})}  \\

{Hyde+2009} & 46410 & 11 & $2.29\pm 0.02$ & & $0.286 \pm 0.002$ \\
Belli+2014 & 56 &11 & $2.38 \pm 0.01$ & & $0.30 \pm 0.03$ \\
Rines+2016 & 123 & 11 & $1.73 \pm 0.03$ & &$0.32 \pm 0.03$ \\
\enddata
\tablecomments{Parameters of the fits. Column 1 identifies the sample and Column 2 gives the number of object in the sample. Column 3 is the ``break point" stellar mass. An error indicates that the break point is fit to the data; otherwise $M_b$ is fixed. Column 4 gives the velocity dispersion at $M_b$. Column 5 and 6 give the power law index of the fit. An entry for $\alpha_2$ alone indicates a single power law fit to the data.}
\label{tab:fit_param}
\end{deluxetable*}

\acknowledgements

HJZ gratefully acknowledges the generous support of the Clay Postdoctoral Fellowship. MJG and DGF are supported by the Smithsonian Institution. We thank Scott Tremaine for carefully reading the manuscript and providing many useful comments. This work benefitted from discussions with Ken Rines, Antonaldo Diaferio, Neta Bahcall, Mark Vogelsberger and Paul Torrey. Masato Onodera kindly pointed out a mistake in Equation 5. This research has made use of NASA's Astrophysics Data System Bibliographic Services. Observations reported here were obtained at the MMT Observatory, a joint facility of the University of Arizona and the Smithsonian Institution. 

Funding for SDSS-III has been provided by the Alfred P. Sloan Foundation, the Participating Institutions, the National Science Foundation, and the U.S. Department of Energy Office of Science. The SDSS-III web site is http://www.sdss3.org/. SDSS-III is managed by the Astrophysical Research Consortium for the Participating Institutions of the SDSS-III Collaboration including the University of Arizona, the Brazilian Participation Group, Brookhaven National Laboratory, University of Cambridge, Carnegie Mellon University, University of Florida, the French Participation Group, the German Participation Group, Harvard University, the Instituto de Astrofisica de Canarias, the Michigan State/Notre Dame/JINA Participation Group, Johns Hopkins University, Lawrence Berkeley National Laboratory, Max Planck Institute for Astrophysics, Max Planck Institute for Extraterrestrial Physics, New Mexico State University, New York University, Ohio State University, Pennsylvania State University, University of Portsmouth, Princeton University, the Spanish Participation Group, University of Tokyo, University of Utah, Vanderbilt University, University of Virginia, University of Washington, and Yale University.

\bibliographystyle{aasjournal}
\bibliography{/Users/jabran/Documents/latex/metallicity}

 \end{document}